\documentclass[useAMS,usenatbib]{mn2e}

\usepackage{graphicx}
\usepackage{dcolumn}
\usepackage{bm}
\usepackage{url}
\newcommand{\PR}[1]{\ensuremath{\left[#1\right]}} 
\newcommand{\PC}[1]{\ensuremath{\left(#1\right)}} 

\title[Constraining Recent Oscillations in Quintessence Models with Euclid]{Constraining Recent Oscillations in Quintessence Models with Euclid}
\author[N. A. Lima and P. T. P. Viana and I. Tereno]{N. A. Lima$^{1,2}$\thanks{E-mail:
ndal@roe.ac.uk (NAL); viana@astro.up.pt (PTPV); tereno@fc.ul.pt (IT)} and P. T. P. Viana$^{2 \star}$ and I. Tereno$^{3 \star}$\\
$^{1}$Institue for Astronomy, University of Edinburgh, Royal Observatory of Edinburgh, Blackford Hill, Edinburgh\\
$^{2}$Centro de Astrof\'{\i}sica, Universidade do Porto, Rua das Estrelas, 4150-762 Porto, Portugal \\
$^{3}$Centro de Astronomia e Astrof\'{i}sica da Universidade de Lisboa, Tapada da Ajuda, 1349-018, Lisboa, Portugal}
\begin{document}

\date{in original form 2014 January 11}

\pagerange{\pageref{firstpage}--\pageref{lastpage}} \pubyear{2014}

\maketitle

\label{firstpage}

\begin{abstract}
Euclid is a future space-based mission that will constrain dark energy with unprecedented accuracy. Its photometric component is optimized for Weak Lensing studies, while the spectroscopic component is designed for Baryon Acoustic Oscillations (BAO) analysis. We use the Fisher matrix formalism to make forecasts on two quintessence dark energy models with a dynamical equation of state that leads to late-time oscillations in the expansion rate of the Universe. We find that Weak Lensing will place much stronger constraints than the BAO, being able to discriminate between oscillating models by measuring the relevant parameters to $1\sigma$ precisions of 5 to $20\%$. The tight constraints suggest that Euclid data could identify even quite small late-time oscillations in the expansion rate of the Universe.
\end{abstract}


\section{\label{Int}Introduction}

The luminosity distance-redshift relation obtained using type Ia Supernova provided the first clear evidence for an accelerating Universe (\citealt{accel1,accel2}). Further evidence, based on different types of observational data, has since become available (\citealt{acc}). The acceleration of the expansion rate of the Universe rapidly became one of the most intriguing problems in modern Cosmology. Since then, there have been many solutions proposed and some have survived the increasingly tight constraints imposed by observational data. Among those solutions, which include changes to the left-hand side (or the geometry/gravity part) of Einstein's field equations, as in modified gravity theories (\citealt{fr1,fr2}), or the introduction of extra spatial dimensions, as in braneworld models (\citealt{brane1}), one of the most popular assumes the existence of an exotic form of energy in the Universe, characterized by a negative pressure, known as dark energy. If it exists, then present-day observations suggest it 
accounts for approximately $75$ per cent of the Universe's total energy density, and its pressure-to-density ratio (known as the equation of state, $w$) is nearly constant with time and close to $-1$ (\citealt{de1,de2,de3,de4}; see \citealt{cp2013} for more details). If future observations continue to point towards a value of $w$ near $-1$, then the hypothesis of the dark energy being in the form of a cosmological constant, $\Lambda$, will gain strength, demanding for more insight on the theoretical problems raised by such a scenario (see \citealt{cc1} for a review). But it is also conceivable that observational data will become more easily reproducible with a value for $w$ close to, but not exactly equal to, $-1$, opening the door for alternative hypothesis, such as quintessence theories.

In quintessence theories, the dark energy is assumed to be the result of a scalar field, $\phi$, minimally coupled to ordinary matter through gravity. The field evolves in a potential $V(\phi)$ with a canonical kinetic term in its lagrangian, which results in a dark energy equation of state $w \geq -1$, not necessarily constant. These models have already been extensively studied (\citealt{q1,q2,q3,q4}) and, usually, it is assumed that the evolution of the scalar field is monotonic in a potential that satisfies the typical inflationary slow roll conditions (see \citealt{inflation1} for a review on inflation), as given by:
\begin{equation}{\label{s1}}
\PC{\frac{1}{V}\frac{dV}{d\phi}}^{2} << 1,
\end{equation}
\noindent and
\begin{equation}{\label{s2}}
\PC{\frac{1}{V}\frac{d^{2}V}{d\phi^{2}}} << 1.
\end{equation}
\noindent Under these conditions, the dark energy equation of state will always be close to $-1$, and the motion of the field becomes greatly simplified (\citealt{sr1}), allowing for an analytical expression for $w$ as a function of the scale factor $a$. The motion of the field will also depend on the present-day values of $w$ and of the dark energy density, $\Omega_{\phi}$, where $\Omega_{\phi}$ is the ratio of the dark energy density, $\rho_{\phi}$, to the universe's critical density, $\rho_c$.

However, the inflationary slow-roll conditions are sufficient but not strictly necessary in quintessence models. Relaxing one of the conditions opens up the possibility of complex scalar field dynamics, allowing for a more dynamical evolution of the dark energy equation of state, while still being on average close to $-1$. In this paper we will be particularly interested in scalar field models where late-time oscillations in the expansion rate of the Universe can arise due to non-standard scalar field dynamics. In particular, we will focus on two scalar field models for which such behavior is possible. 

Model I assumes a quadratic potential,  where an extra degree of freedom, related to the curvature of the potential at the extremum, is introduced on the evolution of the scalar field (\citealt{q5,q6,q7}). More specifically, we consider the case where the curvature of the potential is such that it enables the field to oscillate around the stable extremum, a model already confronted with supernovae data (\citealt{q6}). Here we determine the evolution of its equation-of-state and study its impact on structure formation, forecasting the viability of the model against future Euclid data. 

 In the case of Model II, we consider rapid oscillations in the equation-of-state, corresponding to an effectively constant equation of state when averaged over the oscillation period (see \citealt{rapid1}, \citealt{rapid3}, and references therein). An oscillating equation-of-state is a way to solve the coincidence problem, removing the special character of the current accelerating phase, and of including the crossing of the phantom barrier $w=-1$. Various phenomenological parameterizations of the amplitude, frequency and phase of the oscillations have been proposed (\citealt{wosc1,wosc2,wosc3,structure}), not necessarily arising from oscillations in quintessence models, and have been constrained mostly with background observables, except for \cite{structure} where the impact on structure formation have been investigated. In our paper, the rapid oscillations are produced by a scalar field in a power-law potential where the minimum is zero. The field starts evolving monotonically, and only close to today it 
starts oscillating around the minimum, guaranteeing that the observed cosmic history is not 
significantly affected (\citealt{rapid4}), and producing rapid oscillations in the equation-of-state.

In section 2, we review the dynamics of quintessence and present the models on which we focus our attention. In section 3, we present the method and cosmological probes we will use to obtain the Euclid forecasts presented in section 4. We conclude in section 5.  Throughout the paper we shall assume the metric signature $\PR{-,+,+,+}$ and natural units of $c = \hbar = 8 \pi G = 1$.  


\section{\label{quintessence}QUINTESSENCE MODELS}

The action for quintessence (see \citealt{quintreview} for a thorough review) is given by
\begin{equation}
S = \int d^{4}x\sqrt{-g}\PC{-\frac{1}{4}R + \mathcal{L}_{m} + \mathcal{L}_{\phi}},
\end{equation}
\noindent where $R$ is the Ricci scalar, $\mathcal{L}_{m}$ and $\mathcal{L}_{\phi}$ are, respectively, the matter and scalar field lagrangians. The field's lagrangian takes the form 
\begin{equation}
\mathcal{L}_{\phi} = -\frac{1}{2}g^{\mu \nu}\PC{\partial_{\mu} \phi}\PC{\partial_{\nu} \phi} - V\PC{\phi},
\end{equation}
\noindent where $V(\phi)$ is the field's potential. 

We will be considering the line element for a flat, homogeneous and isotropic Universe as given by the Friedmann-Roberston-Walker (FRW) metric. Varying the field's individual action with respect to the metric elements $g^{\mu \nu}$, one can obtain the field's energy-momentum tensor elements, $T_{\mu \nu}\PC{\phi}$. From the latter, the quintessence equation of state, defined as the ratio between the dark energy pressure, $p_{\phi}$, and energy density, $\rho_{\phi}$, is given by 







\begin{equation}{\label{eqstate}}
w = \frac{p_\phi}{\rho_\phi} = \frac{(1/2)\dot{\phi}^{2} - V(\phi)}{(1/2)\dot{\phi}^{2} + V(\phi)}.
\end{equation}
\noindent This equation shows that, in order to have a dark energy equation of state close to $-1$, the field's evolution has to be potential dominated, such that $\dot{\phi}^{2} << V(\phi)$. This is why, generically, quintessence models are associated to scalar fields slowly evolving in the respective potentials. This can be ensured, for instance, by the slow-roll conditions given by equations (\ref{s1}) and (\ref{s2}).

The scalar field equation of motion is given by the Euler-Lagrange equation 
\begin{equation}{\label{eqmotion}}
\ddot{\phi} + 3H\dot{\phi} + \frac{dV}{d\phi} = 0.
\end{equation}
\noindent Through this equation we can see that the field will evolve in the potential $V(\phi)$ rolling towards a minimum in the quintessence potential, while its motion is damped by the presence of the Hubble parameter, $H$. Considering a flat Universe, with the FRW metric, the Hubble parameter, as a function of the scale factor $a$, is given by $H^{2} = \rho_{_T}/3$, where $\rho_{_T}$ is the Universe's total energy density. We will be considering a Universe consisting of pressureless matter and a scalar field playing the role of dark energy. This implies that $\rho_{_T} = \rho_{m} + \rho_{\phi}$. As a function of the scale factor, $\rho_{m} = \rho_{m0}a^{-3}$, where $\rho_{m0}$ is the present-day value of the matter density; the field's energy density will be given by $\rho_\phi$. In a flat Universe, the present-day value of the matter energy density is determined by $\Omega_{m0} = 1 - \Omega_{\phi0}$, where $\Omega_{m0}$ and $\Omega_{\phi0}$ are the ratios of the present-day values of the matter and dark 
energy densities to the critical density, $\rho_c$.



\subsection{\label{m1}Model I}

The first model we consider is that of a scalar field rolling close to a stable non-zero minimum of its potential. The field is assumed to evolve in a potential that satisfies the slow-roll condition of equation (\ref{s1}), while the other slow-roll condition is somewhat relaxed since $(1/V)d^{2}V/d\phi^{2}$ can be large, namely as a result of the curvature at the potential's minimum. This model was extensively studied by \cite{q6}, where a set of solutions for the evolution of the field was derived, as well as the respective equation of state, $w(a)$, as a function of the scale factor $a$. These solutions are applicable to a wide array of potentials with non-zero minima, and depend on the present-day values of $w$ and $\Omega_{\phi}$, and on the curvature of the potential at the minimum, controlled by the parameter $K^{2}$. The latter is given by equation $(19)$ in \cite{q6}, which we reproduce here:
\begin{equation}{\label{ksquared}}
K^{2} = 1 - \frac{4}{3}\frac{V^{\prime\prime}(\phi_{\star})}{V(\phi_{\star})},
\end{equation}
\noindent where $V^{\prime \prime}(\phi_{\star})$, each prime representing a derivative with respect to the scalar field, and $V(\phi_{\star})$ are, respectively, the curvature and potential values at the minimum of the potential, $\phi_{\star}$. The latter corresponds to the present-day value of the dark energy density, $V(\phi_\star) = \rho_{\phi0}$. For this model we have $V^{\prime \prime}(\phi_{\star})>0$, which means that $-\infty \leq K^{2} \leq 1$. 


In the case of $K^{2} < 0$, the behavior of $w(a)$ is oscillatory, and can be analytically approximated through a combination of sinusoidal functions (\citealt{q6}). For this to happen, we need to have $V^{\prime \prime}(\phi_\star)/V(\phi_\star) > 3/4$, requiring the potential to be significantly curved at its minimum.
In our analysis we do not use the analytical approximation for $w(a)$ derived by \cite{q6}, but instead calculate $w$ directly from equation (\ref{eqstate}) by numerically solving equation (\ref{eqmotion}) as a function of $a$. The solution depends on three parameters: $\Omega_{\phi0}$, $K^{2}$ and the initial value for the field, $\phi_i$. The field's initial velocity, $\dot{\phi}$ is assumed to be zero, which  determines the initial value of the equation of state to be $-1$.

\begin{figure}
\center
    \includegraphics[scale = 0.485]{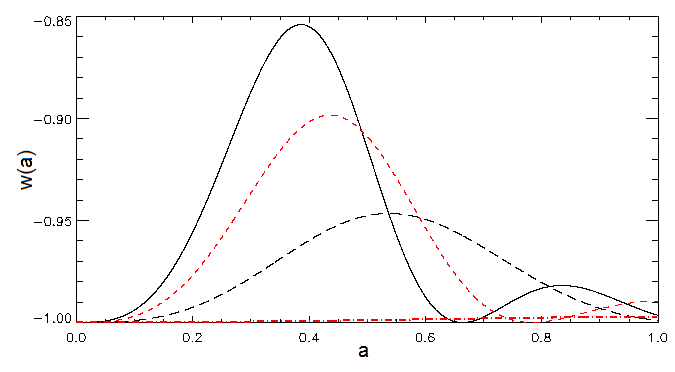}
    \caption{\label{w1} Evolution of $w$ for Model I, assuming different values of $K^{2}$: $K^{2} = 0$ (dot-dashed red line); $K^{2} = -10$ (long dashed black line); $K^{2} = -20$ (short dashed red line), and $K^{2} = -30$ (solid black line). The values of $\Omega_{\phi_0}$ and $\phi_i$ were fixed at $0.74$ and $0.30$, respectively.}
\end{figure}

We assume a quadratic potential, such that
\begin{equation}{\label{pot1}}
V(\phi) = \rho_{\phi_0} + V_{2} \phi^{2},
\end{equation}
\noindent where $V_{2}$ is the curvature at the minimum, determined by $V_{2} = (3/8)\PC{1-K^{2}}\rho_{\phi0}$.
For models with large negative values of the curvature parameter $K^{2}$, the curvature of the potential at the minimum is larger, and the field is able to significantly overcome the Hubble friction term and cross the potential's minimum, oscillating around it with decreasing amplitude. This produces a more dynamical behavior for $w$, as shown in Fig. \ref{w1}, exhibiting damped oscillations with local maxima located between points where the field's motion comes to a halt. For lower absolute values of $K^{2}$, $w$ exhibits a less dynamical behaviour, evolving monotonically in the case $K^{2}=0$. We note that if $\phi_i = 0$,no oscillations occur, regardless the values of the other parameters,and the field effectively behaves as a cosmological constant.

\subsection{\label{m2}Model II}

The second model we consider is that of a scalar field oscillating around zero, with a potential of the form (\citealt{rapid4})
\begin{equation}{\label{pot2}}
V(\phi) = \frac{m^{2}M^{2}}{2}\PR{\frac{\PC{\phi/M}^{2}}{1 + \PC{\phi/M}^{2(1-\alpha)}}}.
\end{equation}
\noindent This potential is very flat for small field values and acquires a quadratic form for large field values. The scale $m$ determines the curvature of the potential at the minimum, $V^{\prime \prime}(0)$, whereas the scale $M$ determines where the potential changes shape and enters the quadratic region, where the field will eventually oscillate. 

This model allows, if certain conditions are met, to have a field slowly-rolling until recently and then present some oscillatory behavior close to the present. It has been show that for a power-law potential, $V(\phi) \propto |\phi|^{n}$, one obtains $w = (n-2)(n+2)$ when averaged over the oscillation period, $T$, as long as it is much smaller than the Hubble time, i.e., $T << H^{-1}$ (\citealt{rapid1,rapid3}). Therefore, in the oscillatory region ($\phi << M$) of the potential being considered, $w$ should average to zero, because we then effectively have $V(\phi) \propto |\phi|^{2}$, i.e. $n=2$. This means the scalar field would behave like pressureless matter, which goes against the need for having a negative pressure component to drive the cosmic acceleration. 

\begin{figure}
\center
    \includegraphics[scale = 0.49]{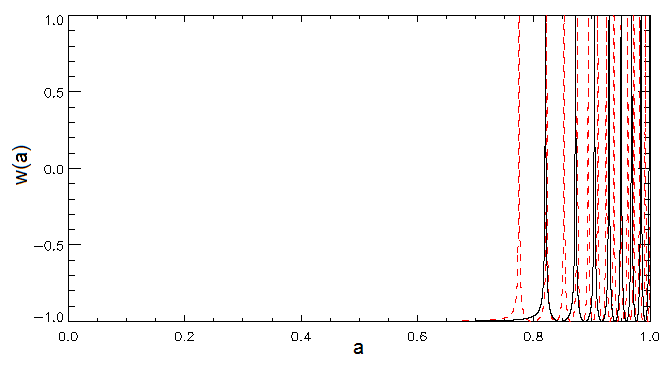}
    \caption{\label{w2} Evolution of $w$ for Model II, assuming different values of $M$: $M = 0.0018$ (dashed red line), and $M = 0.0022$ (solid black line). The value of  $\Omega_{\phi_0}$ was  assumed to be $0.75$.}
\end{figure}

However, this model can be fine-tuned so that the oscillatory region of the potential is reached close enough to the present in order for the predicted large-scale dynamics of the Universe not deviate much from that expected in a model where it is the presence of a cosmological constant that makes the Universe accelerate. This fine-tuning can be achieved if the parameters of the model being considered take values around $\Omega_{\phi_0}=0.75$, $m/H_0 = 1130.6$, with $H_0$ being the present-day value of the Hubble parameter, $M=0.002$ and $\phi_i /M = 23.7$, where $\phi_i $ is the initial value of the field (\citealt{rapid4}). In this particular realization of Model II, the oscillatory behavior only starts around $a = 0.8$, avoiding the appearance of the gravitational instabilities that \cite{unstable} have shown to be a problem for early rapidly oscillating models with a negative averaged equation of state. We add that the frequency of oscillations in $w(a)$ is much larger in model II than in model I, as 
shown in Fig. \ref{w2}. We note that a model with a higher $M$ enters the quadratic region of the potential sooner and thus the oscillations would start earlier, while an increase of  $\phi_i$ has the opposite effect. However, in Fig. \ref{w2} the model with higher $M$ is the one where the oscillations start later. This steems from the fact that we keep the ratio $\phi_i /M$ fixed in our analysis, and shows that the effect of $\phi_i$ is the dominating one.



\section{\label{euclid}Methodology}

Euclid \footnote{http://www.euclid-ec.org/} is a space mission scheduled to be launched by the European Space Agency in $2020$. Its main objective is to help better constrain the large-scale geometry of the Universe and nature of the dark energy and dark matter components of the Universe. In particular, Euclid aims to measure the dark energy parameters $w_0$ and $w_a$ with $2 \%$ and $10 \%$ accuracy, respectively (\citealt{euclidred}), and will be able to constrain a large variety of dark energy models (\citealt{euclidfund}).
To achieve its core scientific objectives, Euclid will use the information contained in the Baryonic Acoustic Oscillations (BAO) present in the matter power spectrum and in the Weak Lensing (WL) features of the large-scale distribution of matter, acquiring spectroscopic and photometric data.


\subsection{\label{bao}Baryonic Acoustic Oscillations}

The formation of acoustic waves in the primordial photon-baryon plasma, in the early stages of our Universe, is imprinted in the form of a series of peaks in the cosmic microwave background radiation (CMB) and in the large-scale distribution of matter (BAO). It presents a characteristic scale, the sound horizon at recombination, $s$, which can be accurately measured using present CMB data, such as that acquired by WMAP (\citealt{wmap}). Its identification, both in the transverse, $y$, and radial, $y^{\prime}$, directions using the galaxy power spectrum, then allows the comoving distance, $\chi(z)$, and the Hubble parameter, $H(z)$, as a function of redshift, to be constrained, given that (\citealt{baorev})
as 
\begin{equation}
y = \frac{\chi(z)}{s}
\end{equation}
\noindent and
\begin{equation}
y^{\prime} = \frac{c}{H(z)s}.
\end{equation}

In order to constrain the two quintessence models with Euclid's spectroscopic survey, we will use the iCosmo software package (\citealt{icosmo}). We first compute $y$ and $y^{\prime}$ given the models' equations-of-state and then compute the Fisher matrix. The inverse of the Fisher matrix contains lower limits for the errors of the estimated values for the models' parameters, according to the Cramer-Rao theorem (\citealt{fisher1}). The BAO Fisher matrix combines the information on $y$ and $y^{\prime}$ and is given by (\citealt{fisherbao}) 
\begin{equation}
\begin{array}{clr}
F_{\alpha \beta}^{BAO} &=& \sum_i \frac{1}{y\PC{z_i}^{2}x_{i}^{2}}\frac{\partial y\PC{z_i}}{\partial p^{\alpha}} \frac{\partial y\PC{z_i}}{\partial p^{\beta}} + \\
&+& \sum_i \frac{1}{y^{\prime}\PC{z_i}^{2}x^{\prime 2}_{i}} \frac{\partial y^{\prime}\PC{z_i}}{\partial p^{\alpha}} \frac{\partial y^{\prime}\PC{z_i}}{\partial p^{\beta}},
\end{array}
\end{equation}
\noindent Here the sums run over the observational bins at different redshifts $z$, $p^{\alpha}$ represents the model parameter with respect to which the partial derivative is taken, while $x$ and $x^{\prime}$ represent the relative errors associated to the measurements of the transverse and radial scales, respectively. We note that iCosmo uses the analytical approximation of \cite{blake} to evaluate the expected accuracy of an experiment's measurement of the BAO scales. We assumed the following Euclid's spectroscopic wide survey specifications (\citealt{euclidred}): the galaxy redshift distribution of \cite{geach}, for a limiting flux of $3 \times 10^{-6}$ erg s$^{-1}$cm$^{-2}$; a median redshift of $1.1$; a redshift measurement error of $\Delta z = 0.001(1+z)$; and a redshift range of $[0.4, 2.2]$ distributed over $14$ redshift bins. 


\subsection{\label{wl}Weak Lensing}

 Weak Lensing is a subtle gravitational lensing effect, caused by the slight deflection of the light emitted by distant galaxies due to the existing matter distribution along their line of sight. The mapping between source, $(x_{2},y_{2})$, and image, $(x_{1},y_{1})$, planes can be approximated to first order through a simple matrix transformation (\citealt{wl1,wl2})
\begin{equation}
  \left(\begin{array}{c}
         x_{2}\\
      y_{2} \end{array}\right) = \left(\begin{array}{cc}
		-\kappa - \gamma_{1} & -\gamma_{2}\\
	   -\gamma_{2} & -\kappa + \gamma_{1} \end{array} \right) \left(\begin{array}{c}
									  x_{1}\\
					  y_{1} \end{array}\right),
\end{equation}
\noindent where $\gamma_{1}$ and $\gamma_{2}$ are the components of the anisotropic shear and $\kappa$ is the isotropic convergence. The magnitude of the lensing signal, measured in shear correlation functions, depends both on the amount of matter along the line of sight, and on the distances between the observer, the lens and the source. This makes weak lensing ideal for measuring the Universe's mass distribution and geometry and, therefore, for constraining cosmological parameters, such as those that can be used to characterize the dark energy.


 We use the iCosmo software to forecast the WL constraints on our models. We evaluate two-point correlation functions of the shear field on separate photometric redshift bins. The Fisher matrix associated with this so-called weak lensing tomography is given by 
\begin{equation}
F_{\alpha \beta}^{WL} = \sum_{l}\frac{2l + 1}{2}f_{sky}\frac{\partial C_{ij}(l)}{\partial p^{\alpha}}\PR{C_{l}}^{-1}_{jk}\frac{\partial C_{km}(l)}{\partial p^{\beta}},
\end{equation}
\noindent where  $C_{ij}(l)$ is the cross-power spectrum between bins $i$ and $j$ for the multipole $\ell$. Here $p^{\alpha}$ represents the model parameter with respect to which the partial derivative is calculated, $f_{sky}$ is the fraction of the sky covered by the survey, and $\PR{C_{l}}_{ij}$ is the covariance matrix for a given $l$ mode and the $i-j$ bin pair (\citealt{icosmo,wlt3}), which depend on the specifications of the Euclid survey. Following \cite{euclidred}, we use a number density of galaxies of $30$ per arcmin$^{2}$, a sky coverage of $15.000$ deg$^2$ and a galaxy redshift distribution given by the analytical formula provided by \cite{smail}, with a median redshift of $0.9$ and parameters $\alpha = 2$ and $\beta = 1.5$ (\citealt{wlt2,wlt1}). The photometric redshift measurement error is assumed to be $\Delta z / (1 + z) = 0.05$, while the random mean-square intrinsic shear per component, $\gamma_{int}$, was taken to be $0.22$ (\citealt{wlt2}). 



Assuming that the Universe contains only photons, baryons, cold dark matter and a quintessence scalar field, as long as this last component does not dominate the total energy density of the Universe at early times, namely before recombination, the transfer function of \cite{icosmotransfer} can be used to determine the power spectrum of the initial matter density perturbations. We use thus this transfer function and evaluate the linear growth function for the quintessence equations-of-state of our models. In order to infer the non-linear corrections to the evolving linear power spectrum, we used the HALOFIT model (\citealt{smithhalofit}), which may lead to some loss of precision when applied to dark energy models with evolving $w$ (\citealt{macdonald}). We compute auto and cross power spectra for $10$ redshift bins between $0 < z < 2$ and for a multipole range limited to $\ell_{\rm max}=5000$ to discard effects of baryonic feedback on the power spectrum (\citealt{semboloni}). 


\section{\label{results}RESULTS}

In this section, we present the results for Euclid's expected constraints on the parameters of the quintessence models under consideration. Given that there aren't a priori theoretically preferred values for these parameters, we will set their fiducial values to those that maximize the joint posterior probability distribution of the parameters given the most recent data available from the Supernova Cosmology Project \footnote{http://supernova.lbs.gov/} (SCP).

\subsection{\label{fiducial}Fiducial models}

Supernovae type Ia are standard candles, since they reach essentially the same luminosity at their peak brightness. The ratio of this constant peak luminosity to the measured flux is proportional to the luminosity distance squared. If we also know the redshift at which a supernova occurs, we can then determine the relation between co-moving distance and redshift, which carries information on all the cosmological parameters that affect it. Namely, those that characterize the quintessence models we are considering.

Assuming that, in each of the quintessence models being considered, all possible values for the free parameters have equal prior probability, then the joint posterior probability distribution of the parameters for each model is proportional to the likelihood of the most recent SCP data. This is provided as
\begin{equation}{\label{distanmod1}}
\mu(z) \equiv m - M,
\end{equation}
where $z$, $m$ and $M$ are, respectively, the redshift, (measured) apparent and (assumed) absolute magnitudes of the observed supernovae. The quantity $\mu(z)$ is related to the cosmology-dependent luminosity distance,  $d_{L}$ (in parsecs), through
\begin{equation}{\label{distanmod2}}
\mu(z) = 5 \log_{10}d_{L}-5.
\end{equation}

Given that the uncertainty associated with the SCP estimates of  $\mu(z)$ is assumed to be well characterized through a Gaussian probability distribution, the likelihood of the most recent SCP data is then proportional to the exponential of $-\chi^{2}/2$, where (\citealt{statistics})
\begin{equation}{\label{chisquaredscp}}
\chi^{2} = \sum_{i,j}[f_i(p) - \mu_i(z_i)]\PR{E^{-1}}_{ij}[f_j(p) - \mu_j(z_i)].
\end{equation}
\noindent with the $\mu_i(z_i)$ and $E^{-1}$ representing, respectively, the SCP data and the inverse of its covariance matrix, which includes systematic uncertainties. The $f_i(z_i,p)$ are the theoretically expected values for the $\mu_i$, assuming the measured redshift $z_i$ and the set of parameters $p$ for the model $f$ to be true.

 For the Model I, we have allowed $\Omega_{\phi0}$, $\phi_i$ and $K^{2}$ to vary freely within the intervals $[0.70,0.80]$, $[0.0,1.0]$ and $[-50.0,0.0]$, respectively. In the case of Model II, we have allowed $\Omega_{\phi0}$ and $M$ to vary freely within the intervals $[0.60,0.80]$ and $[0.00180,0.00220]$, respectively. The interval allowed for $M$, as well as the conditions $\phi_i  = 23.7\,M$ and $m = 1130.6\,H_0^{-1}$, were set to force the scalar field to only start oscillating around $a = 0.8$, rendering the model viable as discussed before. The present-day value of the Hubble parameter was always assumed to be $H_0 = 70\;{\rm km}\,{\rm s}^{-1}\,{\rm Mpc}^{-1}$. For each model, the combination of parameters that maximizes the likelihood of the most recent SCP data, and hence the joint posterior given our assumptions of flat priors for the parameters, is
\begin{itemize}
\item Model I:  $\Omega_{\phi0}=0.74$, $K^{2}=-15.0$, $\phi_i=0.30$
\item Model II: $\Omega_{\phi0}=0.75$, $M=0.00202$
\end{itemize}

The $\chi^2$ values of these two models that best-fit the SCP data are $\chi^2_I=530.63$ and $\chi^2_{II}=535.56$. For comparison, we fitted the same data with $\Lambda CDM$ models with similar number of free parameters, and found a value $\chi^2=531.01$ for its best-fit model. The values are comparable, with the best-fit of model I fitting slghtly better the data than any $\Lambda CDM$ model. 

In the subsequent analysis, we take these best-fit values as our fiducial model parameter sets. We will also always assume to be working on a spatially-flat Universe, thus $\Omega_{m0}=1-\Omega_{\phi0}$, where the power spectrum of primordial adiabatic Gaussian matter density perturbations is approximately scale-invariant $n_{\rm s} = 1$, the abundance of baryons is taken to be $0.045$ and the present-day amplitude of the matter density perturbations smoothed on a sphere with a radius of $8\;{\rm h}^{-1}\,{\rm Mpc}$ is equal to $\sigma_8 = 0.8$, motivated by the WMAP-9 results (\citealt{wmap}).

\begin{figure}
\center
   \includegraphics[scale = 0.38]{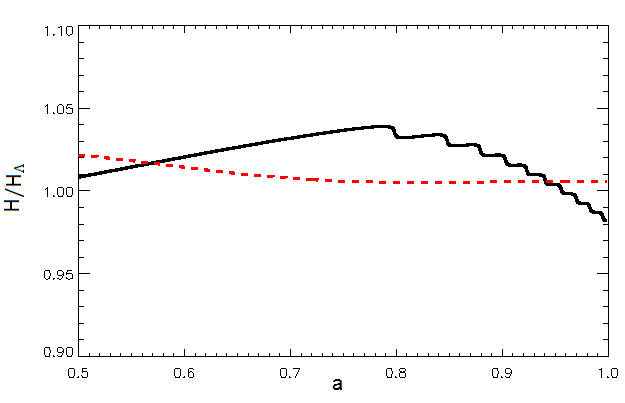}
   \caption{\label{h} Recent evolution of the ratio between the Hubble Parameter, $H(a)$, as a function of the scale factor $a$, evaluated for the fiducial parameters of both models and the corresponding $\Lambda CDM$ (with $\Omega_{\Lambda}=\Omega_{\phi0}$).  The black (full) and red (dashed) lines correspond to Models I and II, respectively.}
\end{figure}

In Fig. \ref{h} we show the evolution of the Hubble parameter, $H(a)$, for the two fiducial models, normalized by $H(a)$ in $\Lambda CDM$. The deviation evolves with redshift and is within $5\%$.

\subsection{\label{confidence} Confidence Regions for Euclid}

\begin{figure}
\begin{center}$
\begin{array}{c}
\includegraphics[scale = 0.42]{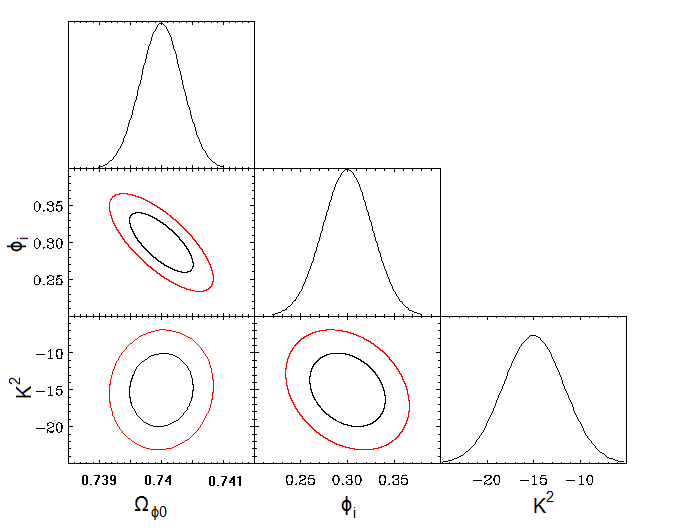} \\
\includegraphics[scale = 0.42]{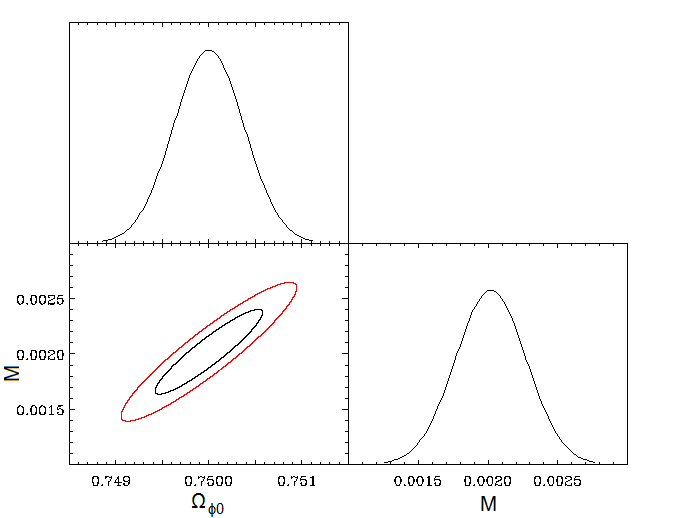}
\end{array}$
\end{center}
\caption{\label{constraints1}It is shown the $68 \%$ (red line) and $95 \%$ (black line) forecasted confidence regions obtained when the Weak Lensing method is applied to Euclid's photometric survey. The upper panel refers to model I, and the lower panel to model II. For both models, we have assumed as central values the fiducial values determined in the previous section. We also plot the one-dimensional distributions for each parameter.}
\end{figure}

\begin{figure}
\begin{center}$
\begin{array}{c}
\includegraphics[scale = 0.42]{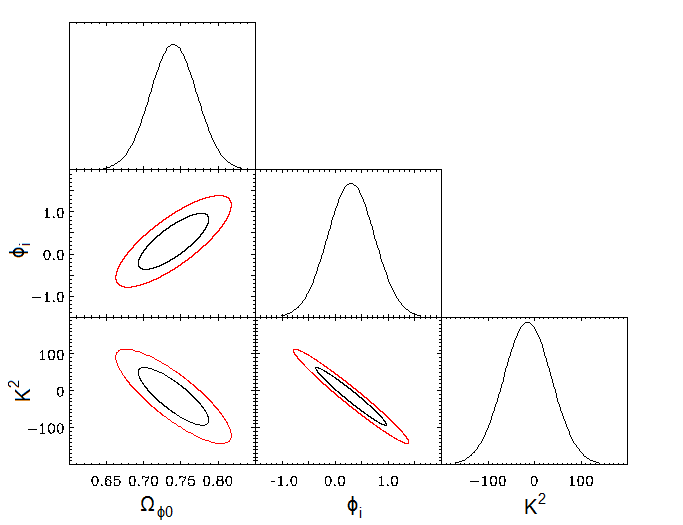} \\
\includegraphics[scale = 0.42]{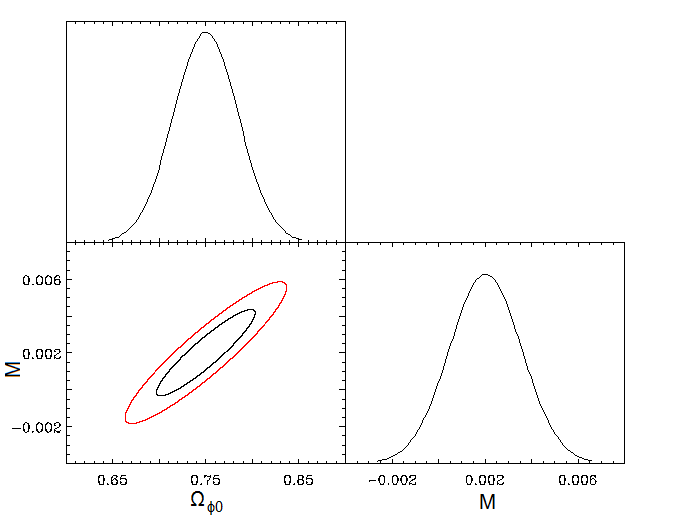}
\end{array}$
\end{center}
\caption{\label{constraints2}It is shown the $68 \%$ (red line) and $95 \%$ (black line) forecasted confidence regions obtained when the BAO method is applied to Euclid's spectroscopic survey. The upper panel refers to model I, and the lower panel to model II. For both models, we have assumed as central values the fiducial values determined in the previous section. We also plot the one-dimensional distributions for each parameter.}
\end{figure}

Applying the Fisher matrix formalism, we compute the marginalized constraints on the model parameters for the Euclid forecasts. Two-dimensional $68 \%$ and $95 \%$ confidence regions, alongside with the one dimensional distributions are shown in Fig. \ref{constraints1} and Fig. \ref{constraints2} for WL and BAO, respectively.


There is a striking difference between the WL and BAO results, with cosmic shear being an order of magnitude more constraining, and dominating the joint constraints.  Note, however, that the full information contained in the spectroscopically-derived matter power spectrum, $P(k)$, such as its full shape, amplitude and redshift distortions, that will be available from Euclid (\citealt{scaramella}), was not used in the analysis, and thus galaxy clustering data may still prvide complementary constraints for the models considered.

We find that the core model parameters, i.e, K, which determines the potential of Model I, and M, which sets the starting redshift of the oscillations in Model II, will be constrained by Euclid with around $15\%$ $1\sigma$ precision.    


 In the case of Model I, we observe an anti-correlation between $K^{2}$ and $\phi_{i}$. This is expected, since a more negative $K^{2}$ would lead to earlier oscillations in the quintessence field, which could be prevented with higher values of $\phi_{i}$. On the other hand, WL and BAO are complementary in the $(\Omega_{\phi0}, K^{2})$ plane, showing a correlation and an anti-correlation, respectively. 

 In the case of Model II, there is a positive correlation between the two parameters. In general, the oscillations in the field start sooner both for larger $M$ and $\Omega_{\phi0}$. However, fixing the ratio $\phi_i /M$, to keep the start of the oscillatory period roughly constant, produces the opposite behaviour as discussed earlier on. Hence, an increase of $M$ (and the corresponding decrease of $\phi_i$) are compensated by a larger $\Omega_{\phi0}$ producing thus a positive correlation.


Tables \ref{table1} and \ref{table2} summarize the $1\sigma$ constraints for Models I and II, respectively.


    \begin{table}
    \begin{center}
    {
    \begin{tabular}[c]{| c | c | c | c |}\hline
     & Weak Lensing & BAO & Combined\\ \hline
    $\Omega_{\phi0}$ & $0.000337$ & $0.0312$ & $0.000332$\\ \hline
    $K^{2}$ & $3.281$ & $51.734$ & $3.190$ \\ \hline
    $\phi_i$ & $0.0267$ & $0.438$ & $0.0262$ \\ \hline
    \end{tabular}}
    \caption{\label{table1} The $1\sigma$ marginalized values for the first model parameters obtained by applying the Weak Lensing and BAO methods, individually and jointly, to Euclid's wide survey}
    \end{center}
    \end{table}

    \begin{table}
    \begin{center}
    {
    \begin{tabular}[c]{| c | c | c | c |}\hline
     & Weak Lensing & BAO & Combined\\ \hline
    $\Omega_{\phi0}$ & $0.000380$ & $0.0350$ & $0.000358$\\ \hline
    $M$ & $0.000253$ & $0.00155$ & $0.000236$ \\ \hline
    \end{tabular}}
    \caption{\label{table2} The $1\sigma$ marginalized values for the second model parameters obtained by applying the Weak Lensing and BAO methods, individually and jointly, to Euclid's wide survey}
    \end{center}
    \end{table}

\section{Conclusions}

In this paper we considered two physically-motivated quintessence models and forecasted their parameter constraints for the future Euclid space mission. The first model has a non-monotonical evolution of the dark energy equation-of-state, while in the second one the equation-of-state oscillates rapidly in the late universe, in contrast to the current paradigm of $\Lambda CDM$ for which the equation of state takes a constant value of $-1$ throughout cosmic evolution. 

In order to obtain plausible fiducial values for the parameters of the models, we have used the most recent data available from the Supernova Cosmology Project. 
For both cases we obtained fiducial values consistent with an overall expansion history of the Universe very similar to that of $\Lambda CDM$, although the recent evolution of the equation-of-state for both models is very different than the $\Lambda CDM$ one. Moreover, the fiducial oscillating models were found to be as good a fit to the data as the concordance $\Lambda CDM$ model, in agreement with the results of \cite{wosc3}, where significant statistical evidence was found for a variety of phenomenological oscillating models (depending on the information criterium used). 

This encouraging result motivated us to pursue the analysis and test our models using probes of structure. Hence, using a Fisher matrix approach, we considered Euclid core probes, weak lensing and galaxy clustering (BAO only). The impact of the dark energy oscillations on the cosmic shear signal is subtle, even with tomography, since oscillations as function of redshift are smeared out by integrations over $w(z)$. Nevertheless, we forecast Euclid's cosmic shear data to have the power to discriminate them from the the concordance $\Lambda CDM$ model. This is in agreement with \cite{structure}, where analysing phenomenological models, weak lensing was considered the best approach to discriminate dark energy oscillations.


\section*{Acknowledgments}

We would like to thank Henk Hoekstra and Martin Kunz for their comments on this manuscript. 

NAL, PTPV and IT acknowledge financial support from Funda\c{c}\~{a}o para a Ci\^{e}ncia e a Tecnologia (FCT), respectively, through grant SFRH/BD/85164/2012, project PTDC/FIS/111725/2009, and grant SFRH/BPD/65122/2009. IT acknowledges further FCT support from Project CERN/FP/123618/2011 and PEst-OE/FIS/UI2751/2014.

\end{document}